\documentclass[12pt]{article}
\usepackage{amssymb}
\usepackage{amsmath}
\usepackage{graphicx}
\usepackage{indentfirst}
\usepackage{cite}
\usepackage{appendix}

\allowdisplaybreaks

\begin{document}

\title{Cross sections for inelastic $K$+$\phi$ scattering}
\author{Yi-Hao Pan and Xiao-Ming Xu$^1$}
\date{}
\maketitle \vspace{-1cm}
\centerline{$^1$Department of Physics, Shanghai University, Baoshan,
Shanghai 200444, China}

\begin{abstract}
In the first Born approximation we study the reactions $K\phi\to\pi K$, 
$\rho K$, $\pi K^*$, and $\rho K^*$ with quark-antiquark annihilation and
creation. Transition amplitudes are derived with the development in spherical
harmonics of the relative-motion wave functions of the two initial mesons and
of the two final mesons so that parity is conserved and the total angular
momentum of the final mesons equals the one of the initial mesons. 
Unpolarized cross sections are calculated from the transition amplitudes that
also contain mesonic quark-antiquark relative-motion wave functions and 
transition potentials for quark-antiquark annihilation and creation. Notable 
temperature dependence of the cross sections is shown. While the cross sections
for $K\phi\to\rho K$, $K\phi\to\pi K^*$, and $K\phi\to\rho K^*$ may be of the
millibarn scale, the cross section for $K\phi\to\pi K$ is very small.
\end{abstract}

\noindent
Keywords: meson-meson scattering, quark-antiquark annihilation and creation, 
quark potential model.

\noindent
PACS: 25.75.-q; 24.85.+p; 12.38.Mh

\clearpage
\vspace{0.5cm}
\leftline{\bf I. INTRODUCTION}
\vspace{0.5cm}

Since enhanced $\phi$ yield was suggested as a signature for the 
formation of quark-gluon plasmas \cite{RM,AS}, many measurements on $\phi$ 
mesons have been made in relativistic heavy-ion collisions such as Au-Au 
collisions at the BNL Relativistic Heavy Ion Collider 
\cite{STAR2007,STAR2009PRC,STAR2009PLB,PHENIX2011,STAR2013,STAR2016,STAR2020,
STAR2022,PHENIX2023} 
or Pb-Pb collisions at the CERN Large Hadron Collider 
\cite{ALICE2015,ALICE2017,ALICE2020,ALICE2022}. Measured ratios like 
$\phi/\pi$, $\phi/K$, and $\Omega/\phi$ show
enhancement of $\phi$ mesons produced in relativistic heavy-ion collisions 
relative to $p$+$p$ collisions. This indicates that strange quarks and strange 
antiquarks are produced in parton-parton scattering in initial nucleus-nucleus
collisions and deconfined matter. Combination of a strange quark and a strange 
antiquark forms a $\phi$ meson at hadronization of the quark-gluon plasma.
The $\phi$ meson in collisions with hadrons in hadronic matter may be broken,
and this changes the $\phi$ yield. For example, the $\phi$ nuclear 
modification factor as a function of transverse momentum is smaller than 1 for
central and midcentral Au-Au collisions at the center-of-mass energy per
nucleon-nucleon pair $\sqrt{s_{NN}}=200$ GeV \cite{STAR2007,PHENIX2011} and
for central and midcentral Pb-Pb collisions at $\sqrt{s_{NN}}=2.76$ TeV 
\cite{ALICE2017}.
Therefore, studying inelastic hadron-$\phi$ scattering is a fundamental issue
in relativistic heavy-ion collisions.

Hadron-$\phi$ reactions can be studied in hadron degrees of freedom
\cite{BR,KS,Haglin,SH,CLK,AK,MKAN} or quark degrees of freedom \cite{LX,XW}.
Starting from an effective meson Lagrangian, Feynman diagrams with one-kaon
exchange were considered, and squared invariant amplitudes for 
$\pi \phi \to K\bar{K}^* +K^*\bar{K}$, $\rho \phi \to K\bar K$, and 
$\phi \phi \to K\bar K$ were provided in Ref. \cite{KS}. 
Using a Lagrangian that is based on an effective
theory in which vector mesons are identified as the dynamical gauge bosons
of the hidden $U(3)_V$ local symmetry in the $U(3)_L \times U(3)_R/U(3)_V$
nonlinear sigma model, large cross sections for $K^* \phi \to \pi K$ and
$K\phi \to \pi K^*$ are shown in Ref. \cite{MKAN}. With a $\pi \phi \rho$
coupling cross sections for $\phi N \to \pi N$, $\phi N \to \rho N$, and
$\phi N \to \pi \Delta$ were obtained in Ref. \cite{CLK}. With a 
$N\Lambda K$ coupling cross sections for $\phi N \to K\Lambda$ are shown to be
much larger than those for $\phi N \to \pi N$, $\phi N \to \rho N$, and
$\phi N \to \pi \Delta$. Experimental efforts to extract the $\phi +N$ total 
cross section from $d(\gamma,pK^+K^-)n$ have been made by the CLAS 
Collaboration \cite{CLAS}. The $\phi +N$ cross section may lead to a
difference in $\phi$ production between $\pi^-$-induced reactions on C
and W targets \cite{HADES}. Inelastic $\pi+\phi$ scattering and inelastic 
$\rho+\phi$ scattering were studied in Ref. \cite{LX} in the quark interchange 
mechanism \cite{BS,Swanson}. 
Adopting temperature dependence in a quark potential, mesonic quark-antiquark
wave functions, and meson masses, prominent temperature-dependent cross 
sections for the inelastic $\pi+\phi$ and $\rho+\phi$ scattering in hadronic 
matter were obtained \cite{LX}. Besides pions and rho mesons, kaons in 
hadronic matter also interact with $\phi$
mesons. However, quark-level study of inelastic $K+\phi$ scattering has not 
been done. Moreover, temperature dependence of the inelastic $K+\phi$ 
scattering
is unexplored both experimentally and theoretically. Therefore, the present
work aims to study the inelastic $K+\phi$ scattering and their temperature
dependence.

Some meson-meson reactions may be dominated by the process where a quark and an
antiquark annihilate into a gluon and subsequently the gluon creates another 
quark-antiquark pair. Such quark-antiquark annihilation and creation has been
used in Refs. \cite{SXW,WX} to obtain unpolarized cross sections for 
the reactions:
$\pi \pi \to \rho \rho$, $K \bar
{K} \to K^* \bar {K}^\ast$, $K \bar{K}^\ast \to K^* \bar{K}^\ast$, $K^\ast
\bar{K} \to K^* \bar{K}^\ast$, $\pi \pi \to K \bar K$, $\pi \rho \to
K \bar {K}^\ast$, $\pi \rho \to K^* \bar{K}$, $K \bar {K} \to \rho \rho$,
$K \bar {K} \to K \bar {K}^\ast,~ K \bar{K} \to K^* \bar{K},
~\pi K \to \pi K^\ast,~ \pi K \to \rho K,
~\pi \pi \to K \bar{K}^\ast,~ \pi \pi \to K^\ast \bar{K},
~\pi \pi \to K^\ast \bar{K}^\ast, ~\pi \rho \to K \bar{K},
~\pi \rho \to K^\ast \bar{K}^\ast, ~\rho \rho \to K^\ast \bar{K}^\ast,
~K \bar{K}^\ast \to \rho \rho$, and $K^* \bar{K} \to \rho \rho$.
The $s$ quark (or the $\bar s$ antiquark) of a kaon may annihilate with the
$\bar s$ antiquark (or the $s$ quark) of a $\phi$ meson to produce a gluon, and
subsequently the gluon splits into a $u\bar u$ or $d\bar d$ pair. The $u\bar u$
or $d\bar d$ pair
combines with spectator constituents of the $K$ and $\phi$
mesons to form two mesons that are not $\phi$ mesons. Quark-antiquark 
annihilation and creation thus leads to inelastic $K+\phi$ scattering. By 
contrast quark interchange does not cause inelastic $K+\phi$ scattering. The 
mechanism that governs the inelastic $K+\phi$ scattering completely differs
from the
mechanism that governs the inelastic $\pi+\phi$ and $\rho+\phi$ scattering.
Therefore, with quark-antiquark annihilation and creation in the first Born 
approximation, in the present work we study the reactions: $K\phi\to\pi K$, 
$K\phi\to\rho K$, $K\phi\to\pi K^*$, and $K\phi\to\rho K^*$. 

This paper is organized as follows. In the next section we derive 
transition-amplitude formulas for 2-to-2 meson-meson scattering that are 
driven by quark-antiquark annihilation and creation. Numerical results and 
relevant discussions are given in Sec. III. A summary is in the last 
section.

\vspace{0.5cm}
\leftline{\bf II. FORMALISM}
\vspace{0.5cm}

The reaction $A(q_1\bar{q}_1)+B(q_2\bar{q}_2) \to
C(q_{3}\bar{q}_{1})+D(q_{2}\bar{q}_{4})$ 
($A(q_1\bar{q}_1)+B(q_2\bar{q}_2) \to 
C(q_{1}\bar{q}_{4})+D(q_{3}\bar{q}_{2})$) 
takes place due to that quark $q_1$ ($q_2$) and antiquark $\bar{q}_2$ 
($\bar{q}_1$) in the initial mesons
annihilate into a gluon and subsequently the gluon creates quark $q_3$ 
and antiquark $\bar{q}_4$. The two processes $q_1+\bar{q}_2 \to q_{3}
+\bar{q}_4$ and $\bar{q}_1+q_2 \to q_{3}+\bar{q}_4$ give rise to
the two transition potentials
$V_{{\rm a}q_1\bar{q}_2}$ and $V_{{\rm a}\bar{q}_1q_2}$, respectively.
Denote by $E_{\rm i}$ and $\vec{P}_{\rm i}$ ($E_{\rm f}$ and $\vec{P}_{\rm f}$)
the total energy and the total momentum of the two initial (final) mesons,
respectively. Let $E_A$ ($E_B$, $E_C$, $E_D$) be the energy of
meson $A$ ($B$, $C$, $D$), and
$V$ the volume where every meson wave function is normalized.
The $S$-matrix element for $A+B \to C+D$ is
\begin{equation}
S_{\rm fi}  = \delta_{\rm fi} -(2\pi)^4 i \delta (E_{\rm f} - E_{\rm i})
\delta^3 (\vec{P}_{\rm f} - \vec{P}_{\rm i})
\frac {{\cal M}_{{\rm a}q_1\bar{q}_2}+{\cal M}_{{\rm a}\bar{q}_1q_2}}
{{V^2}\sqrt{2E_A2E_B2E_C2E_D}},
\end{equation}
where ${\cal M}_{{\rm a}q_1\bar{q}_2}$ and
${\cal M}_{{\rm a}\bar{q}_1q_2}$
are the transition amplitudes given by
\begin{eqnarray}\label{amp_q1qb2}
{\cal M}_{{\rm a}q_1 {\bar q}_2}&=&\frac 
{{(m_{q_3}+m_{\bar{q}_1})}^3}{{m^{3}_{\bar{q}_1}}}\sqrt {2E_A2E_B2E_C2E_D}
\nonumber\\
&&
\times \int d\vec{r}_{q_1\bar{q}_1} d\vec{r}_{q_2\bar{q}_4} 
d\vec{r}_{q_3\bar{q}_1, q_2\bar{q}_4} \psi_{CD}^+ V_{{\rm a}q_1\bar{q}_2}
\nonumber\\
&&
\times \psi_{AB} e^{i\vec{p}_{q_1\bar{q}_1,q_2\bar{q}_2}\cdot
\vec{r}_{q_1\bar{q}_1,q_2\bar{q}_2}-i\vec{p}_{q_3\bar{q}_1, q_2\bar{q}_4}\cdot
\vec{r}_{q_3\bar{q}_1,q_2\bar{q}_4}},
\end{eqnarray}
\begin{eqnarray}\label{amp_qb1q2}
{\cal M}_{{\rm a}\bar{q}_1q_2}&=&\frac 
{{(m_{q_1}+m_{\bar{q}_4})}^3}{m_{q_1}^3}\sqrt {2E_A2E_B2E_C2E_D}
\nonumber\\
&&
\times \int d\vec{r}_{q_1\bar{q}_1} d\vec{r}_{q_3\bar{q}_2} 
d\vec{r}_{q_1\bar{q}_4, q_3\bar{q}_2} \psi_{CD}^+ V_{{\rm a}\bar{q}_1q_2}
\nonumber\\
&&
\times \psi_{AB} e^{i\vec{p}_{q_1\bar{q}_1,q_2\bar{q}_2}\cdot
\vec{r}_{q_1\bar{q}_1,q_2\bar{q}_2}-i\vec{p}_{q_1\bar{q}_4, q_3\bar{q}_2}\cdot
\vec{r}_{q_1\bar{q}_4,q_3\bar{q}_2}},
\end{eqnarray}
where $m_{q_{1}}$ ($m_{\bar{q}_{1}}$, $m_{q_{3}}$, $m_{\bar{q}_{4}}$) is the 
mass of $q_{1}$ ($\bar{q}_{1}$, $q_{3}$, $\bar{q}_{4}$); $\vec {r}_{ab}$ the 
relative coordinate of constituents $a$ and $b$;
$\vec{r}_{q_1\bar {q}_1,q_2\bar {q}_2}$ 
($\vec {r}_{q_3\bar {q}_1,q_2\bar {q}_4}$, 
$\vec {r}_{q_1\bar {q}_4, q_3\bar {q}_2}$) the relative coordinate
of $q_1\bar {q}_1$ and $q_2\bar {q}_2$ ($q_3\bar {q}_1$ and $q_2\bar {q}_4$, 
$q_1\bar {q}_4$ and $q_3\bar {q}_2$);
$\vec {p}_{q_1\bar {q}_1,q_2\bar {q}_2}$ 
($\vec {p}_{q_3\bar {q}_1, q_2\bar {q}_4}$, 
$\vec {p}_{q_1\bar {q}_4, q_3\bar {q}_2}$) the relative momentum 
of $q_1\bar {q}_1$ and $q_2\bar {q}_2$ ($q_3\bar {q}_1$ and $q_2\bar {q}_4$, 
$q_1\bar {q}_4$ and $q_3\bar {q}_2$); $\psi_{AB}$ ($\psi_{CD}$) the wave 
function of mesons $A$ and $B$ ($C$ and $D$), and $\psi_{AB}^+$ 
($\psi_{CD}^+$) the Hermitean conjugate of $\psi_{AB}$ ($\psi_{CD}$).
The wave function of mesons $A$ and $B$ is
\begin{equation}\label{eq_6}
\psi_{AB} =\phi_{A\rm color} \phi_{B\rm color} \phi_{A\rm rel} \phi_{B\rm rel}
\chi_{S_A S_{Az}} \chi_{S_B S_{Bz}} \varphi_{AB\rm flavor},
\end{equation}
and the wave function of mesons $C$ and $D$ is
\begin{equation}\label{teacher_1}
\psi_{CD} =\phi_{C\rm color} \phi_{D\rm color} \phi_{C\rm rel} \phi_{D\rm rel}
\chi_{S_C S_{Cz}} \chi_{S_D S_{Dz}} \varphi_{CD\rm flavor},
\end{equation}
where $S_A$ ($S_B$, $S_C$, $S_D$) is the spin of meson $A$ ($B$, $C$, $D$) with
its
magnetic projection quantum number $S_{Az}$ ($S_{Bz}$, $S_{Cz}$, $S_{Dz}$); 
$\phi_{A\rm color}$
($\phi_{B\rm color}$, $\phi_{C\rm color}$, $\phi_{D\rm color}$),
$\phi_{A\rm rel}$
($\phi_{B\rm rel}$, $\phi_{C\rm rel}$, $\phi_{D\rm rel}$), 
and $\chi_{S_A S_{Az}}$ ($\chi_{S_B S_{Bz}}$, $\chi_{S_C S_{Cz}}$, 
$\chi_{S_D S_{Dz}}$)
are the color wave function, the quark-antiquark relative-motion wave function,
and the spin wave function of meson $A$ ($B$, $C$, $D$), respectively;
$\varphi_{AB\rm flavor}$ ($\varphi_{CD\rm flavor}$) is the flavor wave
function of mesons $A$ and $B$ ($C$ and $D$).

The development in spherical harmonics of the relative-motion wave function
of mesons $A$ and $B$ (aside from a normalization constant) is given by
\begin{eqnarray}\label{teacher_5,6}
e^{i\vec{p}_{q_1\bar{q}_1,q_2\bar{q}_2} \cdot
\vec{r}_{q_1\bar{q}_1,q_2\bar{q}_2}} & = & 4\pi
\sum\limits_{L_{\rm i}=0}^{\infty}
\sum\limits_{M_{\rm i}=-L_{\rm i}}^{L_{\rm i}}
i^{L_{\rm i}} j_{L_{\rm i}} (\mid \vec{p}_{q_1\bar{q}_1,q_2\bar{q}_2} \mid
r_{q_1\bar{q}_1,q_2\bar{q}_2})
\nonumber\\
& & \times Y_{L_{\rm i}M_{\rm i}}^\ast
(\hat{p}_{q_1\bar{q}_1,q_2\bar{q}_2}) Y_{L_{\rm i}M_{\rm i}}
(\hat{r}_{q_1\bar{q}_1,q_2\bar{q}_2}),
\end{eqnarray}
and the development in spherical harmonics of the relative-motion wave function
of mesons $C$ and $D$ leads to
\begin{eqnarray}
e^{-i\vec{p}_{q_3\bar{q}_1,q_2\bar{q}_4} \cdot
\vec{r}_{q_3\bar{q}_1,q_2\bar{q}_4}} & = & 4\pi
\sum\limits_{L_{\rm f}=0}^{\infty}
\sum\limits_{M_{\rm f}=-L_{\rm f}}^{L_{\rm f}}
i^{L_{\rm f}} (-1)^{L_{\rm f}} j_{L_{\rm f}} 
(\mid \vec{p}_{q_3\bar{q}_1,q_2\bar{q}_4} \mid r_{q_3\bar{q}_1,q_2\bar{q}_4})
\nonumber\\
& & \times Y_{L_{\rm f}M_{\rm f}}^\ast (\hat{p}_{q_3\bar{q}_1,q_2\bar{q}_4})
Y_{L_{\rm f}M_{\rm f}} (\hat{r}_{q_3\bar{q}_1,q_2\bar{q}_4}),
\end{eqnarray}
in ${\cal{M}}_{aq_{1}\bar{q}_{2}}$, and
\begin{eqnarray}
e^{-i\vec{p}_{q_1\bar{q}_4,q_3\bar{q}_2} \cdot
\vec{r}_{q_1\bar{q}_4,q_3\bar{q}_2}} & = & 4\pi
\sum\limits_{L_{\rm f}=0}^{\infty}
\sum\limits_{M_{\rm f}=-L_{\rm f}}^{L_{\rm f}}
i^{L_{\rm f}} (-1)^{L_{\rm f}} j_{L_{\rm f}} 
(\mid \vec{p}_{q_1\bar{q}_4,q_3\bar{q}_2} \mid r_{q_1\bar{q}_4,q_3\bar{q}_2})
\nonumber\\
& & \times Y_{L_{\rm f}M_{\rm f}}^\ast (\hat{p}_{q_1\bar{q}_4,q_3\bar{q}_2})
Y_{L_{\rm f}M_{\rm f}} (\hat{r}_{q_1\bar{q}_4,q_3\bar{q}_2}),
\end{eqnarray}
in ${\cal{M}}_{a\bar{q}_{1}q_{2}}$, where $Y_{L_{\rm i}M_{\rm i}}$ 
($Y_{L_{\rm f}M_{\rm f}}$) are the spherical harmonics with
the orbital-angular-momentum quantum number $L_{\rm i}$ ($L_{\rm f}$) and the 
magnetic
projection quantum number $M_{\rm i}$ ($M_{\rm f}$), $j_{L_{\rm i}}$ and
$j_{L_{\rm f}}$ are the spherical Bessel
functions, and $\hat{p}_{q_1\bar{q}_1,q_2\bar{q}_2}$
($\hat{p}_{q_3\bar{q}_1,q_2\bar{q}_4}$, $\hat{p}_{q_1\bar{q}_4,q_3\bar{q}_2}$,
$\hat{r}_{q_1\bar{q}_1,q_2\bar{q}_2}$,
$\hat{r}_{q_3\bar{q}_1,q_2\bar{q}_4}$, $\hat{r}_{q_1\bar{q}_4,q_3\bar{q}_2}$)
denotes the polar angles of $\vec{p}_{q_1\bar{q}_1,q_2\bar{q}_2}$
($\vec{p}_{q_3\bar{q}_1,q_2\bar{q}_4}$, $\vec{p}_{q_1\bar{q}_4,q_3\bar{q}_2}$,
$\vec{r}_{q_1\bar{q}_1,q_2\bar{q}_2}$,
$\vec{r}_{q_3\bar{q}_1,q_2\bar{q}_4}$, $\vec{r}_{q_1\bar{q}_4,q_3\bar{q}_2}$).

Let $\chi_{SS_{z}}$ ($\chi_{S^{\prime}S^{\prime}_{z}}$) stand for the spin wave
function of mesons $A$ and $B$ ($C$ and $D$),
which has the total spin $S$ ($S^{\prime}$) and its $z$ component $S_z$ 
($S^{\prime}_{z}$). The
Clebsch-Gordan coefficients $(S_{A}S_{Az}S_{B}S_{Bz}|SS_{z})$ relate
$\chi_{SS_z}$ to $\chi_{S_AS_{Az}}\chi_{S_BS_{Bz}}$, 
and $(S_{C}S_{Cz}S_{D}S_{Dz}|S^{\prime}S^{\prime}_{z})$ relate
$\chi_{S^{\prime}S^{\prime}_{z}}$ to $\chi_{S_CS_{Cz}}\chi_{S_DS_{Dz}}$:
\begin{eqnarray}
\chi_{S_{A}S_{A_{z}}}\chi_{S_{B}S_{B_{z}}}&=&
\sum^{S_{\rm max}}_{S=S_{\rm min}}\sum^{S}_{S_{z}=-S}
(S_{A}S_{Az}S_{B}S_{Bz}|SS_{z})\chi_{SS_{z}},\\
\chi_{S_{C}S_{C_{z}}} \chi_{S_{D}S_{D_{z}}}
&=&
\sum^{S^{\prime}_{\rm max}}_{S^{\prime}=S^{\prime}_{\rm min}}
\sum^{S^{\prime}}_{S^{\prime}_z=-S^{\prime}}
(S_{C}S_{Cz}S_{D}S_{Dz}|S^{\prime}S^{\prime}_z) \chi_{S^{\prime}S^{\prime}_z},
\end{eqnarray}
where $S_{\rm min}=\mid S_A-S_B \mid$, $S_{\rm max}=S_A+S_B$, 
$S^{\prime}_{\rm min}=\mid S_C-S_D \mid$, and $S^{\prime}_{\rm max}=S_C+S_D$. 
$Y_{L_{\rm i}M_{\rm i}}$ and $\chi_{SS_{z}}$ ($Y_{L_{\rm f}M_{\rm f}}$ and 
$\chi_{S^{\prime}S^{\prime}_z}$) are coupled to the wave function
$\varphi^{\rm in}_{JJ_z}$ ($\varphi^{\rm final}_{J^\prime J_z^\prime}$) which 
has the total angular momentum $J$ ($J^{\prime}$) of mesons
$A$ and $B$ ($C$ and $D$) and its $z$ component $J_z$ ($J^{\prime}_z$),
\begin{eqnarray}
Y_{L_{\rm{i}}M_{\rm{i}}}\chi_{SS_{z}} &=&
\sum^{J_{\rm max}}_{J=J_{\rm min}}\sum^J_{J_z=-J}
(L_{\rm{i}}M_{\rm{i}}SS_{z}|JJ_{z})\varphi^{\rm{in}}_{JJ_{z}},\\
Y_{L_{\rm f}M_{\rm f}}\chi_{S^{\prime}S^{\prime}_z} &=&
\sum^{J^{\prime}_{\rm max}}_{J^{\prime}=J^{\prime}_{\rm min}} 
\sum^{J^{\prime}}_{J^{\prime}_z=-J^{\prime}}
(L_{\rm f}M_{\rm f}S^{\prime}S^{\prime}_z|J^{\prime}J^{\prime}_z) 
{\varphi}^{\rm final}_{J^{\prime}J^{\prime}_z},
\end{eqnarray}
where $J_{\rm min}=\mid L_{\rm i}-S \mid$, $J_{\rm max}=L_{\rm i}+S$, 
$J^{\prime}_{\rm min}=\mid L_{\rm f}-S^{\prime} \mid$, and
$J^{\prime}_{\rm max}=L_{\rm f}+S^{\prime}$.
$(L_{\rm{i}}M_{\rm{i}}SS_{z}|JJ_{z})$ and 
$(L_{\rm{f}}M_{\rm{f}}S^{\prime}S^{\prime}_{z}|J^{\prime}J^{\prime}_{z})$ are 
the Clebsch-Gordan coefficients. It
follows from Eqs. (6)-(12) that the transition amplitude given in 
Eq. (3) becomes
\begin{eqnarray}
{\cal M}_{{\rm a}\bar{q}_{1}q_{2}}&=&
\frac{(m_{q_1}+m_{\bar{q}_4})^3}{{m^3_{q_1}}} 
\sqrt{2E_{A}2E_{B}2E_{C}2E_{D}} (4\pi)^2
\sum^{\infty}_{L_{\rm i}=0} \sum^{L_{\rm i}}_{M_{\rm i}=-L_{\rm i}} 
i^{L_{\rm i}}
Y^*_{L_{\rm i}M_{\rm i}}(\hat{p}_{q_{1}\bar{q}_{1},q_{2}\bar{q}_{2}})
\nonumber\\
&&
\times \sum^{\infty}_{L_{\rm f}=0} \sum^{L_{\rm f}}_{M_{\rm f}=-L_{\rm f}} 
i^{L_{\rm f}} (-1)^{L_{\rm f}}
Y^*_{L_{\rm f}M_{\rm f}}(\hat{p}_{q_{1}\bar{q}_{4},q_{3}\bar{q}_{2}})
{\phi}^{+}_{C{\rm color}} {\phi}^{+}_{D{\rm color}} 
{\varphi}^{+}_{CD{\rm flavor}}
\nonumber\\
&&
\times \int{d\vec{r}_{q_{1}\bar{q}_{1}} d\vec{r}_{q_{3}\bar{q}_{2}} 
d\vec{r}_{q_{1}\bar{q}_{4},q_{3}\bar{q}_{2}}}
{\phi}^{+}_{C{\rm rel}} {\phi}^{+}_{D{\rm rel}} 
\sum_{S^{\prime}S^{\prime}_{z}}
(S_{C}S_{Cz}S_{D}S_{Dz}|S^{\prime}S^{\prime}_{z})
\nonumber\\
&&
\times \sum_{J^{\prime}J^{\prime}_{z}}
(L_{\rm f}M_{\rm f}S^{\prime}S^{\prime}_{z}|J^{\prime}J^{\prime}_{z}) 
\varphi^{\rm final}_{J^{\prime}J^{\prime}_{z}}
V_{{\rm a}q_{1}\bar{q}_{2}} \sum_{SS_{z}}
(S_{A}S_{Az}S_{B}S_{Bz}|SS_{z})
\nonumber\\
&&
\times \sum_{JJ_{z}}(L_{\rm i}M_{\rm i}SS_{z}|JJ_{z}) \varphi^{\rm in}_{JJ_{z}}
{\phi}_{A{\rm rel}} {\phi}_{B{\rm rel}} {\varphi}_{AB{\rm flavor}}
\phi_{A{\rm color}}\phi_{B{\rm color}} 
\nonumber\\
&&
\times j_{L_{\rm i}}(\mid \vec{p}_{q_1\bar{q}_1,q_2\bar{q}_2} 
\mid r_{q_1\bar{q}_1,q_2\bar{q}_2}) j_{L_{\rm f}}
(\mid \vec{p}_{q_1\bar{q}_4,q_3\bar{q}_2} \mid r_{q_1\bar{q}_4,q_3\bar{q}_2}).
\end{eqnarray}
Conservation of the total angular momentum implies that $J$ equals 
$J^{\prime}$ and $J_z$ equals $J^{\prime}_{z}$. This leads to
\begin{eqnarray}
{\cal M}_{{\rm a}\bar{q}_{1}q_{2}}&=&
\frac{(m_{q_1}+m_{\bar{q}_4})^3}{{m^3_{q_1}}} 
\sqrt{2E_{A}2E_{B}2E_{C}2E_{D}} (4\pi)^2
\sum^{\infty}_{L_{\rm i}=0} \sum^{L_{\rm i}}_{M_{\rm i}=-L_{\rm i}} 
i^{L_{\rm i}}
Y^*_{L_{\rm i}M_{\rm i}}(\hat{p}_{q_{1}\bar{q}_{1},q_{2}\bar{q}_{2}})
\nonumber\\
&&
\times \sum^{\infty}_{L_{\rm f}=0} \sum^{L_{\rm f}}_{M_{\rm f}=-L_{\rm f}} 
i^{L_{\rm f}} (-1)^{L_{\rm f}}
Y^*_{L_{\rm f}M_{\rm f}}(\hat{p}_{q_{1}\bar{q}_{4},q_{3}\bar{q}_{2}})
{\phi}^{+}_{C{\rm color}} {\phi}^{+}_{D{\rm color}} 
{\varphi}^{+}_{CD{\rm flavor}}
\nonumber\\
&&
\times \int{d\vec{r}_{q_{1}\bar{q}_{1}} d\vec{r}_{q_{3}\bar{q}_{2}} 
d\vec{r}_{q_{1}\bar{q}_{4},q_{3}\bar{q}_{2}}}
{\phi}^{+}_{C{\rm rel}} {\phi}^{+}_{D{\rm rel}} 
\sum_{S^{\prime}S^{\prime}_{z}}
(S_{C}S_{Cz}S_{D}S_{Dz}|S^{\prime}S^{\prime}_{z})
\nonumber\\
&&
\times \sum_{JJ_{z}}(L_{\rm f}M_{\rm f}S^{\prime}S^{\prime}_{z}|JJ_{z}) 
\varphi^{\rm final}_{JJ_{z}}
V_{{\rm a}\bar{q}_{1}q_{2}} \sum_{SS_{z}}(S_{A}S_{Az}S_{B}S_{Bz}|SS_{z})
\nonumber\\
&&
\times (L_{\rm i}M_{\rm i}SS_{z}|JJ_{z}) \varphi^{\rm in}_{JJ_{z}} 
{\phi}_{A{\rm rel}} {\phi}_{B{\rm rel}} {\varphi}_{AB{\rm flavor}}
\phi_{A{\rm color}}\phi_{B{\rm color}} 
\nonumber\\
&&
\times j_{L_{\rm i}}(\mid \vec{p}_{q_1\bar{q}_1,q_2\bar{q}_2} 
\mid r_{q_1\bar{q}_1,q_2\bar{q}_2}) j_{L_{\rm f}}
(\mid \vec{p}_{q_1\bar{q}_4,q_3\bar{q}_2} \mid r_{q_1\bar{q}_4,q_3\bar{q}_2}).
\end{eqnarray}
Using the relation
\begin{eqnarray}
\varphi^{\rm{in}}_{JJ_{z}} &=&
\sum_{\bar{M}_{\rm i}\bar{S}_{z}}
(L_{\rm{i}}\bar{M}_{\rm{i}}S\bar{S}_{z}|JJ_{z})
Y_{L_{\rm{i}}\bar{M}_{\rm{i}}}\chi_{S\bar{S}_{z}},\\
\varphi^{\rm final}_{JJ_{z}} &=&
\sum_{\bar{M}_{\rm f}\bar{S}^{\prime}_{z}}
(L_{\rm f}\bar{M}_{\rm f}S^{\prime}\bar{S}^{\prime}_{z}|JJ_{z})
Y_{L_{\rm f}\bar{M}_{\rm f}} \chi_{S^{\prime}S^{\prime}_{z}},
\end{eqnarray}
where $(L_{\rm{i}}\bar{M}_{\rm{i}}S\bar{S}_{z}|JJ_{z})$ and 
$(L_{\rm{f}}\bar{M}_{\rm{f}}S^{\prime}\bar{S}^{\prime}_{z}|J_{}J_{z})$ are the
Clebsch-Gordan coefficients, we get
\begin{eqnarray}
{\cal M}_{{\rm a}\bar{q}_{1}q_{2}}&=& 
\frac{(m_{q_1}+m_{\bar{q}_4})^3}{{m^3_{q_1}}} 
\sqrt{2E_{A}2E_{B}2E_{C}2E_{D}} (4\pi)^2
\sum^{\infty}_{L_{\rm i}=0} \sum^{L_{\rm i}}_{M_{\rm i}=-L_{\rm i}} 
i^{L_{\rm i}}
Y^*_{L_{\rm i}M_{\rm i}}(\hat{p}_{q_{1}\bar{q}_{1},q_{2}\bar{q}_{2}})
\nonumber\\
&&
\times \sum^{\infty}_{L_{\rm f}=0} \sum^{L_{\rm f}}_{M_{\rm f}=-L_{\rm f}} 
i^{L_{\rm f}} (-1)^{L_{\rm f}}
Y^*_{L_{\rm f}M_{\rm f}}(\hat{p}_{q_{1}\bar{q}_{4},q_{3}\bar{q}_{2}})
\sum_{S^{\prime}S^{\prime}_{z}}
(S_{C}S_{Cz}S_{D}S_{Dz}|S^{\prime}S^{\prime}_{z})
\nonumber\\
&&
\times \sum_{JJ_{z}}(L_{\rm f}M_{\rm f}S^{\prime}S^{\prime}_{z}|JJ_{z})
\sum_{\bar{M}_{\rm f}\bar{S}^{\prime}_{z}}
(L_{\rm f}\bar{M}_{\rm f}S^{\prime}\bar{S}^{\prime}_{z}|JJ_{z})
\sum_{SS_{z}}(S_{A}S_{Az}S_{B}S_{Bz}|SS_{z}) 
\nonumber\\
&&
\times (L_{\rm i}M_{\rm i}SS_{z}|JJ_{z}) 
\sum_{\bar{M}_{\rm i}\bar{S}_{z}}(L_{\rm i}\bar{M}_{\rm i}S\bar{S}_{z}|JJ_{z})
{\phi}^{+}_{C{\rm color}} {\phi}^{+}_{D{\rm color}} 
{\varphi}^{+}_{CD{\rm flavor}}
\chi^{+}_{S^{\prime}\bar{S}^{\prime}_{z}}
\nonumber\\
&&
\times \int{d\vec{r}_{q_{1}\bar{q}_{1}} d\vec{r}_{q_{3}\bar{q}_{2}} 
d\vec{r}_{q_{1}\bar{q}_{4},q_{3}\bar{q}_{2}}}
j_{L_{\rm f}}(\mid \vec{p}_{q_1\bar{q}_4,q_3\bar{q}_2} 
\mid r_{q_1\bar{q}_4,q_3\bar{q}_2})
Y_{L_{\rm f}\bar{M}_{\rm f}}(\hat{r}_{q_{1}\bar{q}_{4},q_{3}\bar{q}_{2}})
\nonumber\\
&&
\times {\phi}^{+}_{C{\rm rel}} {\phi}^{+}_{D{\rm rel}} 
V_{{\rm a}\bar{q}_{1}q_{2}}\phi_{A{\rm rel}}\phi_{B{\rm rel}} 
j_{L_{\rm i}}(\mid \vec{p}_{q_1\bar{q}_1,q_2\bar{q}_2} 
\mid r_{q_1\bar{q}_1,q_2\bar{q}_2})%
Y_{L_{\rm i}\bar{M}_{\rm i}}(\hat{r}_{q_{1}\bar{q}_{1},q_{2}\bar{q}_{2}})
\nonumber\\
&&
\times \chi_{S\bar{S}_{z}} {\varphi}_{AB{\rm flavor}}
\phi_{A{\rm color}}\phi_{B{\rm color}}.
\end{eqnarray}
Furthermore, we need the identity
\begin{equation}
j_l(pr)Y_{lm}(\hat{r})=\int \frac{d^3p^\prime}{(2\pi)^3}\frac{2\pi^2}{p^2}
\delta (p-p^\prime) i^l (-1)^l Y_{lm}(\hat{p}^\prime)
e^{i\vec{p}^{~\prime} \cdot \vec{r}} ,
\end{equation}
which is obtained with the help of
$\int_0^\infty j_l(pr)j_l(p^\prime r)r^2dr=\frac{\pi}{2p^2}
\delta (p-p^\prime)$ \cite{AW,joachain}, and where $\hat{r}$ ($\hat{p}^\prime$)
denotes the polar angles of $\vec r$ ($\vec{p}^{~\prime}$). Substituting Eq. 
(18) in Eq. (17), we get
\begin{eqnarray}
{\cal M}_{{\rm a}\bar{q}_{1}q_{2}}&=& 
\frac{(m_{q_1}+m_{\bar{q}_4})^3}{{m^3_{q_1}}} 
\sqrt{2E_{A}2E_{B}2E_{C}2E_{D}} (4\pi)^2
\sum^{\infty}_{L_{\rm i}=0} \sum^{L_{\rm i}}_{M_{\rm i}=-L_{\rm i}} 
i^{L_{\rm i}}
Y^*_{L_{\rm i}M_{\rm i}}(\hat{p}_{q_{1}\bar{q}_{1},q_{2}\bar{q}_{2}})
\nonumber\\
&&
\times \sum^{\infty}_{L_{\rm f}=0} \sum^{L_{\rm f}}_{M_{\rm f}=-L_{\rm f}} 
i^{L_{\rm f}} (-1)^{L_{\rm f}}
Y^*_{L_{\rm f}M_{\rm f}}(\hat{p}_{q_{1}\bar{q}_{4},q_{3}\bar{q}_{2}})
\sum_{S^{\prime}S^{\prime}_{z}}
(S_{C}S_{Cz}S_{D}S_{Dz}|S^{\prime}S^{\prime}_{z})
\nonumber\\
&&
\times \sum_{JJ_{z}}(L_{\rm f}M_{\rm f}S^{\prime}S^{\prime}_{z}|JJ_{z})
\sum_{\bar{M}_{\rm f}\bar{S}^{\prime}_{z}}
(L_{\rm f}\bar{M}_{\rm f}S^{\prime}\bar{S}^{\prime}_{z}|JJ_{z})
\sum_{SS_{z}}(S_{A}S_{Az}S_{B}S_{Bz}|SS_{z}) 
\nonumber\\
&&
\times (L_{\rm i}M_{\rm i}SS_{z}|JJ_{z}) 
\sum_{\bar{M}_{\rm i}\bar{S}_{z}}(L_{\rm i}\bar{M}_{\rm i}S\bar{S}_{z}|JJ_{z})
{\phi}^{+}_{C{\rm color}} {\phi}^{+}_{D{\rm color}} 
{\varphi}^{+}_{CD{\rm flavor}}
\chi^{+}_{S^{\prime}\bar{S}^{\prime}_{z}}
\nonumber\\
&&
\times \int{\frac{d^{3}p_{\rm frm}}{(2\pi)^{3}}} 
\frac{2\pi^{2}}{\vec{p}^{2}_{{q_{1}}\bar{q}_{4},q_{3}\bar{q}_{2}}}
\delta(\mid \vec{p}_{q_{1}\bar{q}_{4},q_{3}\bar{q}_{2}} 
\mid-\mid \vec{p}_{\rm frm} \mid) 
i^{L_{\rm f}} (-1)^{L_{\rm f}}%
Y_{L_{\rm f}\bar{M}_{\rm f}} (\hat{p}_{\rm frm})
\nonumber\\
&&
\times \int{\frac{d^{3}p_{\rm irm}}{(2\pi)^{3}}} 
\frac{2\pi^{2}}{\vec{p}^{2}_{{q_{1}}\bar{q}_1,q_{2}\bar{q}_{2}}}
\delta(\mid \vec{p}_{q_{1}\bar{q}_1,q_{2}\bar{q}_{2}} 
\mid-\mid \vec{p}_{\rm irm} \mid) i^{L_{\rm i}} (-1)^{L_{\rm i}}%
Y_{L_{\rm i}\bar{M}_{\rm i}} (\hat{p}_{\rm irm})
\nonumber\\
&&
\times \int{d\vec{r}_{q_{1}\bar{q}_{1}} d\vec{r}_{q_{3}\bar{q}_{2}} 
d\vec{r}_{q_{1}\bar{q}_{4},q_{3}\bar{q}_{2}}}
{\phi}^{+}_{C{\rm rel}} {\phi}^{+}_{D{\rm rel}} 
V_{{\rm a}\bar{q}_{1}q_{2}}\phi_{A{\rm rel}}\phi_{B{\rm rel}} 
\nonumber\\
&&
\times e^{i\vec{p}_{\rm frm}\cdot\vec{r}_{q_{1}\bar{q}_4,q_{3}\bar{q}_2}}
e^{i\vec{p}_{\rm irm}\cdot\vec{r}_{q_{1}\bar{q}_1,q_{2}\bar{q}_2}}
\chi_{S\bar{S}_{z}} \varphi_{AB{\rm flavor}}
\phi_{A{\rm color}}\phi_{B{\rm color}}.
\end{eqnarray}
Let $\vec{r}_c$ be the position vector of constituent $c$. $\phi_{A\rm{rel}}$
and $\phi_{B\rm{rel}}$ are functions of
the relative coordinate of the quark and
the antiquark inside mesons $A$ and $B$, respectively. 
We take the Fourier transform of
$V_{{\rm a}q_1\bar{q}_2}$, $V_{{\rm a}\bar{q}_1q_2}$, $\phi_{A\rm{rel}}$,
and $\phi_{B\rm{rel}}$:
\begin{equation}
V_{{\rm a}q_1\bar{q}_2}(\vec{r}_{q_3}-\vec{r}_{q_1}) =
\int \frac {d^3k}{(2\pi)^3} V_{{\rm a}q_1\bar{q}_2} (\vec {k})
e^{i\vec {k} \cdot (\vec{r}_{q_3}-\vec{r}_{q_1})},
\end{equation}
\begin{equation}
V_{{\rm a}\bar{q}_1q_2}(\vec{r}_{q_3}-\vec{r}_{q_2}) =
\int \frac {d^3k}{(2\pi)^3} V_{{\rm a}\bar{q}_1q_2} (\vec {k})
e^{i\vec {k} \cdot (\vec{r}_{q_3}-\vec{r}_{q_2})},
\end{equation}
\begin{equation}
\phi_{A\rm rel}(\vec{r}_{q_1\bar{q}_1}) =
\int \frac {d^3p_{q_1\bar{q}_1}}{(2\pi)^3} \phi_{A\rm rel}
(\vec {p}_{q_1\bar{q}_1})
e^{i\vec {p}_{q_1\bar{q}_1} \cdot \vec {r}_{q_1\bar{q}_1}},
\end{equation}
\begin{equation}
\phi_{B\rm rel}(\vec{r}_{q_2\bar{q}_2}) =
\int \frac {d^3p_{q_2\bar{q}_2}}{(2\pi)^3} \phi_{B\rm rel}
(\vec {p}_{q_2\bar{q}_2})
e^{i\vec {p}_{q_2\bar{q}_2} \cdot \vec {r}_{q_2\bar{q}_2}}.
\end{equation}
In Eqs. (20)-(21) $\vec k$ is the gluon momentum,
and in Eqs. (22)-(23)
$\vec{p}_{ab}$ is the relative momentum of constituents $a$ and $b$. 
In momentum space the normalizations are
\begin{displaymath}
\int \frac{d^3p_{q_1\bar{q}_1}}{(2\pi)^3}
\phi_{A\rm rel}^+(\vec{p}_{q_1\bar{q}_1})
\phi_{A\rm rel}(\vec{p}_{q_1\bar{q}_1})=1,
\end{displaymath}
\begin{displaymath}
\int \frac{d^3p_{q_2\bar{q}_2}}{(2\pi)^3}
\phi_{B\rm rel}^+(\vec{p}_{q_2\bar{q}_2})
\phi_{B\rm rel}(\vec{p}_{q_2\bar{q}_2})=1.
\end{displaymath}
The spherical polar coordinates of $\vec{p}_{\rm irm}$ and $\vec{p}_{\rm frm}$ 
are expressed as
$(\mid \vec{p}_{\rm irm} \mid, \theta_{\rm irm}, \phi_{\rm irm})$ and 
$(\mid \vec{p}_{\rm frm} \mid, \theta_{\rm frm}, \phi_{\rm frm})$, 
respectively. Integration over
$\mid \vec{p}_{\rm irm} \mid$, $\mid \vec{p}_{\rm frm} \mid$, 
$\vec{r}_{q_1\bar{q}_1}$, $\vec{r}_{q_3\bar{q}_2}$, and 
$\vec{r}_{q_1\bar{q}_4,q_3\bar{q}_2}$
in Eq. (19) yields
\begin{eqnarray}
{\cal M}_{{\rm a}\bar{q}_{1}q_{2}}&=&
\sqrt{2E_{A}2E_{B}2E_{C}2E_{D}}
\sum^{\infty}_{L_{\rm i}=0} \sum^{L_{\rm i}}_{M_{\rm i}=-L_{\rm i}}
Y^*_{L_{\rm i}M_{\rm i}}(\hat{p}_{q_{1}\bar{q}_{1},q_{2}\bar{q}_{2}})
\nonumber\\
&&
\times \sum^{\infty}_{L_{\rm f}=0} 
\sum^{L_{\rm f}}_{M_{\rm f}=-L_{\rm f}} (-1)^{L_{\rm f}}
Y^*_{L_{\rm f}M_{\rm f}}(\hat{p}_{q_{1}\bar{q}_{4},q_{3}\bar{q}_{2}})
\sum_{S^{\prime}S^{\prime}_{z}}
(S_{C}S_{Cz}S_{D}S_{Dz}|S^{\prime}S^{\prime}_{z})
\nonumber\\
&&
\times \sum_{JJ_{z}}(L_{\rm f}M_{\rm f}S^{\prime}S^{\prime}_{z}|JJ_{z})
\sum_{\bar{M}_{\rm f}\bar{S}^{\prime}_{z}}
(L_{\rm f}\bar{M}_{\rm f}S^{\prime}\bar{S}^{\prime}_{z}|JJ_{z})
\sum_{SS_{z}}(S_{A}S_{Az}S_{B}S_{Bz}|SS_{z}) 
\nonumber\\
&&
\times (L_{\rm i}M_{\rm i}SS_{z}|JJ_{z}) 
\sum_{\bar{M}_{\rm i}\bar{S}_{z}}(L_{\rm i}\bar{M}_{\rm i}S\bar{S}_{z}|JJ_{z})
{\phi}^{+}_{C{\rm color}} {\phi}^{+}_{D{\rm color}} 
{\varphi}^{+}_{CD{\rm flavor}}
\chi^{+}_{S^{\prime}\bar{S}^{\prime}_{z}}
\nonumber\\
&&
\times \int{d\theta_{\rm frm}d\phi_{\rm frm}} \sin{\theta}_{\rm frm} 
Y_{L_{\rm f}\bar{M}_{\rm f}}(\hat{p}_{\rm frm})
\int{d\theta_{\rm irm}d\phi_{\rm irm}} \sin{\theta}_{\rm irm} 
Y_{L_{\rm i}\bar{M}_{\rm i}}(\hat{p}_{\rm irm})
\nonumber\\
&&
\times \int{\frac{d^{3}p_{q_{1}\bar{q}_{1}}}{(2\pi)^3}}
\int{\frac{d^{3}p_{q_{2}\bar{q}_{2}}}{(2\pi)^3}}
{\phi}^{+}_{C{\rm rel}} (\vec{p}_{q_{1}\bar{q}_{1}}
+\frac{m_{q_1}}{m_{q_{1}}+m_{\bar{q}_{1}}}\vec{p}_{\rm irm}
+\frac{m_{q_1}}{m_{q_{1}}+m_{\bar{q}_{4}}}\vec{p}_{\rm frm})
\nonumber\\
&&
\times {\phi}^{+}_{D{\rm rel}} (\vec{p}_{q_{2}\bar{q}_{2}}
+\frac{m_{\bar{q}_{2}}}{m_{q_{2}}+m_{\bar{q}_{2}}}\vec{p}_{\rm irm}
+\frac{m_{\bar{q}_{2}}}{m_{q_{3}}+m_{\bar{q}_{2}}}\vec{p}_{\rm frm})
\nonumber\\
&&
\times V_{{\rm a}\bar{q}_{1}q_{2}}
[\vec{p}_{q_{2}\bar{q}_{2}}-\vec{p}_{q_{1}\bar{q}_{1}}
-(\frac{m_{q_2}}{m_{q_{2}}+m_{\bar{q}_{2}}}
-\frac{m_{\bar{q}_{1}}}{m_{q_{1}}+m_{\bar{q}_{1}}})\vec{p}_{\rm irm}]
\nonumber\\
&&
\times \phi_{A{\rm rel}}(\vec{p}_{q_{1}\bar{q}_{1}})
\phi_{B{\rm rel}}(\vec{p}_{q_{2}\bar{q}_{2}})
\chi_{S\bar{S}_{z}} \varphi_{AB{\rm flavor}}
\phi_{A{\rm color}}\phi_{B{\rm color}},
\end{eqnarray}
in which $\mid \vec{p}_{\rm irm} \mid = \mid
\vec{p}_{q_1\bar{q}_1,q_2\bar{q}_2} \mid$ and $\mid \vec{p}_{\rm frm} \mid = 
\mid \vec{p}_{q_1\bar{q}_4,q_3\bar{q}_2} \mid$; $\hat{p}_{\rm irm}$
($\hat{p}_{\rm frm}$) denotes the polar angles of $\vec{p}_{\rm irm}$ 
($\vec{p}_{\rm frm}$); $m_{q_2}$ and 
$m_{\bar{q}_2}$ are the $q_2$ and $\bar{q}_2$ masses, respectively.
The expression of the other transition amplitude 
${\cal M}_{{\rm a}q_1\bar{q}_2}$ is similar to the right-hand side in Eq.
(24), and is thus given from ${\cal M}_{{\rm a}\bar{q}_1q_2}$ by replacing
$\hat{p}_{q_{1}\bar{q}_{4},q_{3}\bar{q}_{2}}$
($\vec{p}_{q_{1}\bar{q}_{1}}
+\frac{m_{q_1}}{m_{q_{1}}+m_{\bar{q}_{1}}}\vec{p}_{\rm irm}
+\frac{m_{q_1}}{m_{q_{1}}+m_{\bar{q}_{4}}}\vec{p}_{\rm frm}$,
$\vec{p}_{q_{2}\bar{q}_{2}}
+\frac{m_{\bar{q}_{2}}}{m_{q_{2}}+m_{\bar{q}_{2}}}\vec{p}_{\rm irm}
+\frac{m_{\bar{q}_{2}}}{m_{q_{3}}+m_{\bar{q}_{2}}}\vec{p}_{\rm frm}$,
$\vec{p}_{q_{2}\bar{q}_{2}}-\vec{p}_{q_{1}\bar{q}_{1}}
-(\frac{m_{q_2}}{m_{q_{2}}+m_{\bar{q}_{2}}}
-\frac{m_{\bar{q}_{1}}}{m_{q_{1}}+m_{\bar{q}_{1}}})\vec{p}_{\rm irm}$)
with 
$\hat{p}_{q_{3}\bar{q}_{1},q_{2}\bar{q}_{4}}$ 
($\vec{p}_{q_{1}\bar{q}_{1}}
-\frac{m_{\bar{q}_1}}{m_{q_{1}}+m_{\bar{q}_{1}}}\vec{p}_{\rm irm}
-\frac{m_{\bar{q}_1}}{m_{q_{3}}+m_{\bar{q}_{1}}}\vec{p}_{\rm frm}$,
$\vec{p}_{q_{2}\bar{q}_{2}}
-\frac{m_{q_{2}}}{m_{q_{2}}+m_{\bar{q}_{2}}}\vec{p}_{\rm irm}
-\frac{m_{q_{2}}}{m_{q_{2}}+m_{\bar{q}_{4}}}\vec{p}_{\rm frm}$,
$\vec{p}_{q_{1}\bar{q}_{1}}-\vec{p}_{q_{2}\bar{q}_{2}}
+(\frac{m_{q_1}}{m_{q_{1}}+m_{\bar{q}_{1}}}
-\frac{m_{\bar{q}_{2}}}{m_{q_{2}}+m_{\bar{q}_{2}}})\vec{p}_{\rm irm}$).
So far, we have obtained new
expressions of the transition amplitudes from Eqs. (2) and (3).

With the transition amplitudes the unpolarized cross section for 
$A+B \to C+D$ is
\begin{eqnarray}
\sigma^{\rm unpol}(\sqrt{s},T) &=&
\frac{1}{(2J_{A}+1)(2J_{B}+1)}\frac{1}{32\pi s}
\frac{\mid \vec{P}^{~\prime}(\sqrt{s}) \mid}{\mid \vec{P}(\sqrt{s}) \mid} 
\nonumber \\
&& 
\times{\int^{\pi}_{0} d\theta \sum_{J_{A_{z}}J_{B_{z}}J_{C_{z}}J_{D_{z}}} 
{\mid {\cal M}_{{\rm a}q_1\bar{q}_2} +{\cal M}_{{\rm a}\bar{q}_1q_2} \mid}^{2}
\sin{\theta}},\label{eq:unpolcs}
\end{eqnarray}
where $s$ is the Mandelstam variable obtained from the four-momenta $P_A$ and
$P_B$ of mesons $A$ and $B$ by $s=(P_A+P_B)^2$; $T$ is the temperature;
$J_A$ ($J_B$, $J_C$, $J_D$) and $J_{Az}$ ($J_{Bz}$, $J_{Cz}$, $J_{Dz}$) 
of meson $A$ ($B$, $C$, $D$) are the total angular momentum and its $z$ 
component, respectively; $\theta$ is the angle between $\vec{P}$ and $\vec{P}'$
which are the three-dimensional momenta of mesons $A$ and $C$ in the 
center-of-mass frame, respectively. 
Let $m_A$, $m_B$, $m_C$, and $m_D$ be the masses of mesons $A$, $B$, $C$,
and $D$, respectively. $\vec P$ and $\vec{P}^\prime$ are given by
\begin{eqnarray}
{\vec{P}}^2(\sqrt{s})&=&\frac{1}{4s} \left[ \left(s-m^2_A-m^2_B \right) ^2
-4m^2_Am^2_B \right],
\\
{\vec{P}}^{\prime 2}(\sqrt{s})&=&\frac{1}{4s}\left[ \left(s-m^2_C-m^2_D 
\right) ^2-4m^2_Cm^2_D \right].
\end{eqnarray}
On the basis of the relativistic energy-momentum relation, we have
\begin{equation}
E_A=\sqrt{\vec{P}^2+m_A^2}=\frac{1}{2\sqrt s}(s+m_A^2-m_B^2),
\end{equation}
\begin{equation}
E_B=\sqrt{\vec{P}^2+m_B^2}=\frac{1}{2\sqrt s}(s-m_A^2+m_B^2),
\end{equation}
\begin{equation}
E_C=\sqrt{\vec{P}^{\prime 2}+m_C^2}=\frac{1}{2\sqrt s}(s+m_C^2-m_D^2),
\end{equation}
\begin{equation}
E_D=\sqrt{\vec{P}^{\prime 2}+m_D^2}=\frac{1}{2\sqrt s}(s-m_C^2+m_D^2).
\end{equation}
We calculate the cross section in the center-of-mass frame of the two initial 
mesons.
According to the Feynman rules, the two processes $q_1+\bar{q}_2 \to q_{3}
+\bar{q}_4$ and $\bar{q}_1+q_2 \to q_{3}+\bar{q}_4$ contribute to meson-meson
scattering on an equal footing, and the sum 
${\cal M}_{{\rm a}q_1\bar{q}_2}+{\cal M}_{{\rm a}\bar{q}_1q_2}$ appears in Eq.
(25) if ${\cal M}_{{\rm a}q_1\bar{q}_2} \neq 0$ and 
${\cal M}_{{\rm a}\bar{q}_1q_2} \neq 0$.

\vspace{0.5cm}
\leftline{\bf III. NUMERICAL CROSS SECTIONS AND DISCUSSIONS }
\vspace{0.5cm}

The quark-antiquark relative-motion wave functions, $\phi_{A\rm rel}$ and
$\phi_{B\rm rel}$ in Eq. (4) as well as $\phi_{C\rm rel}$ and
$\phi_{D\rm rel}$ in Eq. (5), are solutions of the Schr\"odinger equation with
a temperature-dependent quark potential. The potential between constituents
$a$ and $b$ in coordinate space is \cite{SXW}
\begin{eqnarray}
V_{ab}(\vec{r}_{ab}) & = &
- \frac {\vec{\lambda}_a}{2} \cdot \frac {\vec{\lambda}_b}{2}
\xi_1 \left[ 1.3- \left( \frac {T}{T_{\rm c}} \right)^4 \right] \tanh 
(\xi_2 r_{ab}) + \frac {\vec{\lambda}_a}{2} \cdot \frac {\vec{\lambda}_b}{2}
\frac {6\pi}{25} \frac {v(\lambda r_{ab})}{r_{ab}} \exp (-\xi_3 r_{ab})~~~~
\nonumber  \\
& & -\frac {\vec{\lambda}_a}{2} \cdot \frac {\vec{\lambda}_b}{2}
\frac {16\pi^2}{25}\frac{d^3}{\pi^{3/2}}\exp(-d^2r^2_{ab}) 
\frac {\vec {s}_a \cdot \vec {s}_b} {m_am_b}
+\frac {\vec{\lambda}_a}{2} \cdot \frac {\vec{\lambda}_b}{2}\frac {4\pi}{25}
\frac {1} {r_{ab}} \frac {d^2v(\lambda r_{ab})}{dr_{ab}^2} 
\frac {\vec {s}_a \cdot \vec {s}_b}{m_am_b},
\end{eqnarray}
where $\xi_1=0.525$ GeV, $\xi_2=1.5[0.75+0.25 (T/{T_{\rm c}})^{10}]^6$ GeV, 
$\xi_3=0.6$ GeV, and
$\lambda=\sqrt{25/16\pi^2 \alpha'}$ with $\alpha'=1.04$ GeV$^{-2}$;
$T_{\rm c}=0.175$ GeV is the critical temperature at which the phase transition
between the quark-gluon plasma and hadronic matter takes place 
\cite{KLP,DPS,XWB};
$m_a$, $\vec{s}_a$, and $\vec{\lambda}_a$ are the mass, the spin, and the
Gell-Mann matrices for the color generators of constituent $a$, respectively;
the dimensionless function $v$ is given by Buchm\"uller and Tye in Ref. 
\cite{BT}; the quantity $d$ is related to constituent quark masses by
\begin{eqnarray}
d^2=d_1^2\left[\frac{1}{2}+\frac{1}{2}\left(\frac{4m_a m_b}{(m_a+m_b)^2}
\right)^4\right]+d_2^2\left(\frac{2m_am_b}{m_a+m_b}\right)^2,
\end{eqnarray}
where $d_1=0.15$ GeV and $d_2=0.705$.
The potential originates from perturbative quantum chromodynamics (QCD) 
at short distances and lattice QCD at intermediate and long distances. 
The first and second terms are the central spin-independent potential of which
the short-distance part arises from one-gluon exchange plus perturbative one- 
and two-loop corrections in vacuum \cite{BT} and the intermediate-distance and 
long-distance part fits well the numerical potential which was obtained in the
lattice gauge calculations \cite{KLP}. The third term is the smeared spin-spin
interaction that comes from one-gluon exchange between constituents $a$ and 
$b$ \cite{GI}, and the fourth term is the spin-spin interaction that arises
from perturbative one- and two-loop corrections to one-gluon exchange 
\cite{Xu2002}. Temperature dependence of the potential is given by the first
term, and comes from the lattice gauge calculations \cite{KLP}. At long
distances the spin-independent potential is independent of $r_{ab}$, and 
obviously exhibits a plateau at $T/T_{\rm c}>0.55$.
The plateau height decreases with increasing temperature. Confinement thus
becomes weaker and weaker.

The Schr\"odinger equation with the potential yields energy eigenvalues and 
quark-antiquark relative-motion wave functions in coordinate space. The sum
of the quark mass, the antiquark mass, and an energy eigenvalue gives a meson 
mass.
In the present work we use the constituent quark masses, 0.32 GeV for the up 
and down quarks and 0.5 GeV for the strange quark. The quark masses are
independent of temperature. The experimental masses of $\pi$, $\rho$, $K$, 
$K^*$, $\eta$, $\omega$, and $\phi$ mesons are reproduced from the
Schr\"odinger equation with the potential at $T=0$. Furthermore, the 
temperature dependence of the potential leads to temperature dependence of
meson masses and mesonic quark-antiquark relative-motion wave functions. The
temperature dependence of the $\pi$, $\rho$, $K$, and $K^*$ masses are shown in
Ref. \cite{SX}, where the temperature covers the temperature region of hadronic
matter, and parametrizations of these meson masses are given. The temperature
dependence of the $\phi$ mass is shown in Ref. \cite{LXW}, and is parametrized
as
\begin{equation}
m_{\phi}=0.931\left[ 1-\left( \frac{T}{1.12T_{\rm c}} \right)^{5.46}
\right]^{1.32}.
\end{equation}
Since confinement becomes weaker and weaker with increasing temperature,
spatial extension of the mesonic quark-antiquark relative-motion wave functions
becomes larger and larger. Since the orbital-angular-momentum quantum numbers
of $\pi$, $\rho$, $K$, $K^*$, $\eta$, $\omega$, and $\phi$ mesons are zero, 
the wave 
functions are not zero at $r_{ab}=0$. When the temperature increases, the 
absolute values of the wave functions at $r_{ab}=0$ decrease.

The transition potentials
$V_{{\rm a}q_1\bar{q}_2}$ and $V_{{\rm a}\bar{q}_1q_2}$ are derived from
perturbative QCD in Ref.
\cite{SXW}. From the wave functions and the transition potentials we get the
transition amplitudes ${\cal M}_{{\rm a}q_{1}\bar{q}_{2}}$ and 
${\cal M}_{{\rm a}\bar{q}_{1}q_{2}}$. In practical calculations the summations
over $L_{\rm i}$ and $L_{\rm f}$ in the transition amplitudes are from 0 to 3.
The orbital-angular-momentum quantum numbers $L_{\rm i}$ and $L_{\rm f}$
are selected to satisfy that parity is conserved and that the total angular
momentum of the two final mesons equals the total angular momentum of the two 
initial mesons. Values of $L_{\rm i}$ and $L_{\rm f}$ are listed in Table 1.

\begin{table*}[htbp]
\caption{\label{table1} Total spin and orbital-angular-momentum quantum 
number.}
\tabcolsep=5pt
\begin{tabular}{ccccc}
  \hline
  \hline
reaction & $S$ & $S^\prime$ & $L_{\rm i}$ & $L_{\rm f}$\\
  \hline
$K\phi \to \pi K$ & 1 & 0 & 1 & 1\\
                  & 1 & 0 & 2 & 2\\
                  & 1 & 0 & 3 & 3\\
  \hline
$K\phi \to \rho K$ & 1 & 1 & 0 & 0,2\\
                   & 1 & 1 & 1 & 1,3\\
                   & 1 & 1 & 2 & 0,2\\
                   & 1 & 1 & 3 & 1,3\\
  \hline
$K\phi \to \pi K^*$ & 1 & 1 & 0 & 0,2\\
                    & 1 & 1 & 1 & 1,3\\
                    & 1 & 1 & 2 & 0,2\\
                    & 1 & 1 & 3 & 1,3\\
  \hline
$K\phi \to \rho K^*$ & 1 & 0 & 1 & 1\\
                     & 1 & 0 & 2 & 2\\
                     & 1 & 0 & 3 & 3\\
                     & 1 & 1 & 0 & 0,2\\
                     & 1 & 1 & 1 & 1,3\\
                     & 1 & 1 & 2 & 0,2\\
                     & 1 & 1 & 3 & 1,3\\
                     & 1 & 2 & 0 & 2\\
                     & 1 & 2 & 1 & 1,3\\
                     & 1 & 2 & 2 & 0,2\\
                     & 1 & 2 & 3 & 1,3\\
  \hline
  \hline
\end{tabular}
\end{table*}

We consider the four $K$+$\phi$ reactions:
$K\phi\to\pi K$, $K\phi\to\rho K$, $K\phi\to\pi K^*$, and $K\phi\to\rho K^*$. 
${\cal M}_{{\rm a}q_1\bar{q}_2}$ and ${\cal M}_{{\rm a}\bar{q}_1q_2}$ are 
proportional to flavor matrix elements. Since
the flavor matrix elements for the $K$+$\phi$ reactions with total isospin
$I = \frac{1}{2}$ are zero for ${\cal M}_{{\rm a}q_1\bar{q}_2}$ and 
$-\frac{\sqrt{6}}{2}$ for ${\cal M}_{{\rm a}\bar{q}_1q_2}$, only the process 
$\bar{q}_1+q_2 \to q_{3}+\bar{q}_4$ contributes to these reactions. The
unpolarized cross section for the four $K$+$\phi$ reactions is
\begin{eqnarray}
\sigma^{\rm unpol}(\sqrt{s},T) &=&
\frac{1}{(2J_{A}+1)(2J_{B}+1)}\frac{1}{32\pi s}
\frac{\mid \vec{P}^{~\prime}(\sqrt{s}) \mid}{\mid \vec{P}(\sqrt{s}) \mid} 
\nonumber \\
&& 
\times{\int^{\pi}_{0} d\theta \sum_{J_{A_{z}}J_{B_{z}}J_{C_{z}}J_{D_{z}}} 
{\mid {\cal M}_{{\rm a}\bar{q}_1q_2} \mid}^{2}
\sin{\theta}}.
\end{eqnarray}

If the sum of the masses of the two initial mesons of a reaction is larger than
the one of the two final mesons, the reaction is exothermic. Even
slowly-moving initial mesons may start the reaction, and a certain amount of 
the masses of the initial mesons are converted into kinetic energies of the
final mesons. If the sum of the masses of the two initial mesons is smaller 
than
that of the two final mesons, the reaction is endothermic. The initial mesons
need kinetic energies to satisfy energy conservation and to start the reaction,
and a certain amount of the kinetic energies are converted into the masses of 
the final mesons.

The reaction $K\phi\to\rho K^*$ is endothermic at $T/T_c=0$ and exothermic at 
$T/T_c=0.65$, 0.75, 0.85, 0.9, and 0.95. The other three reactions are 
exothermic. Cross sections for exothermic reactions are infinite at threshold 
energies. We thus start calculations of the cross sections for exothermic
reactions at $\sqrt{s}=m_K+m_\phi+10^{-4}$ GeV, where $m_K$ and $m_\phi$ are
the masses of the kaon and the $\phi$ meson, respectively.
Numerical unpolarized cross sections for $K\phi\to\pi K$, $K\phi\to\rho K$, 
$K\phi\to\pi K^*$, and $K\phi\to\rho K^*$ are plotted as red solid curves
in Fig. 1 through Fig. 4. Because the quark potential, the meson masses, and
the mesonic quark-antiquark relative-motion wave functions depend on 
temperature, $\mid \vec{P} \mid$, $\mid \vec{P}^{\prime} \mid$, $E_A$, $E_B$,
$E_C$, $E_D$, and ${\cal M}_{{\rm a}\bar{q}_1q_2}$, which are given in Eq. (24)
and Eqs. (26)-(31), depend on temperature. This leads to temperature dependence
of the unpolarized cross sections.

The numerical cross sections for endothermic reactions are parametrized as
\begin{eqnarray}
\sigma^{\rm unpol}(\sqrt {s},T)
&=&a_1 \left( \frac {\sqrt {s} -\sqrt {s_0}} {b_1} \right)^{c_1}
\exp \left[ c_1 \left( 1-\frac {\sqrt {s} -\sqrt {s_0}} {b_1} \right) \right]
\nonumber \\
&&+ a_2 \left( \frac {\sqrt {s} -\sqrt {s_0}} {b_2} \right)^{c_2}
\exp \left[ c_2 \left( 1-\frac {\sqrt {s} -\sqrt {s_0}} {b_2} \right) \right],
\label{enparaEq}
\end{eqnarray}
where $\sqrt{s_0}$ is the threshold energy, and $a_1$, $b_1$, $c_1$, $a_2$,
$b_2$, and $c_2$ are parameters. The numerical cross sections
for exothermic reactions are parametrized as
\begin{eqnarray}
\sigma^{\rm unpol}(\sqrt {s},T)
&=&\frac{\vec{P}^{\prime 2}}{\vec{P}^2}
\left\{a_1 \left( \frac {\sqrt {s} -\sqrt {s_0}} {b_1} \right)^{c_1}
\exp \left[ c_1 \left( 1-\frac {\sqrt {s} -\sqrt {s_0}} {b_1} \right) \right]
\right.
\nonumber \\
&&+ \left.
a_2 \left( \frac {\sqrt {s} -\sqrt {s_0}} {b_2} \right)^{c_2}
\exp \left[ c_2 \left( 1-\frac {\sqrt {s} -\sqrt {s_0}} {b_2} \right) \right]
\right\}.
\label{exparaEq}
\end{eqnarray}
The parameter values are listed in Table 2. $d_0$ is the
separation between the peak's location on the $\sqrt s$-axis and the threshold
energy, and $\sqrt{s_z}$ is the square root of the Mandelstam variable at which
the cross section is 1/100 of the peak cross section. For the endothermic
reaction $K\phi \to \rho K^*$ at $T=0$ a peak is displayed in Fig. 4, and 
$d_0=0.25$ GeV and $\sqrt{s_z}=3.04$ GeV are obtained from the numerical cross 
section for $K\phi \to \rho K^*$.
About exothermic reactions we may not see peak cross sections, but 
$\vec{P}^2/\vec{P}^{~\prime 2}$ times numerical cross sections for exothermic
reactions must show peak cross sections. Hence, for exothermic reactions $d_0$ 
and $\sqrt{s_z}$ are obtained from
$\vec{P}^2/\vec{P}^{~\prime 2}$ times the numerical cross sections.

The cross sections given by the parametrizations are plotted as green dashed
curves in Fig. 1 through Fig. 4. For the exothermic reactions the solid and
dashed curves look coincided. For the endothermic reaction
$K\phi \to \rho K^*$ at $T=0$, difference between the solid curve and the 
dashed curve exists around the two peak cross sections and at $\sqrt {s}>2.2$ 
GeV.

The threshold energy for each of the exothermic reactions 
$K\phi\to\pi K$, $K\phi\to\rho K$, and $K\phi\to\pi K^*$ are the sum of the 
$K$ and $\phi$ masses. When the
temperature increases, decreases of the masses lead to a decrease in the 
threshold energy as shown in Figs. 1-3. As $\sqrt s$ increases 
near the threshold energy, the cross 
sections for these reactions decrease rapidly due to the factor 
$\mid \vec{P}^{~\prime} \mid / \mid \vec{P} \mid$ in Eq. (35).
The threshold energy of the endothermic reaction 
$K\phi\to\rho K^*$ at $T=0$ is the sum of the $\rho$ and $K^*$ masses.
As $\sqrt s$ increases from the threshold energy, the cross section for 
$K\phi\to\rho K^*$ at $T=0$ increases rapidly from zero, reaches a peak value 
of about 1.66 mb, and then decreases.

Since the reactions $K\phi\to\pi K$, $K\phi\to\rho K$, and $K\phi\to\pi K^*$
are exothermic, one always find exclusive final states $\pi K$, $\rho K$, 
and $\pi K^*$ in a $K+ \phi$ reaction. However, one may not find the
exclusive final state $\rho K^*$ in a $K+ \phi$ reaction in vacuum, since the 
reaction $K\phi\to\rho K^*$ is endothermic at $T=0$. If the total energy
$\sqrt s$ of $K$ and $\phi$ mesons in the center-of-mass frame is smaller than
the threshold energy of $K\phi\to\rho K^*$, the reaction does not occur. If
$\sqrt s$ is larger than the threshold energy, one can observe $\rho K^*$
production. 

\begin{figure}[htbp]
\centering
\includegraphics[scale=0.65]{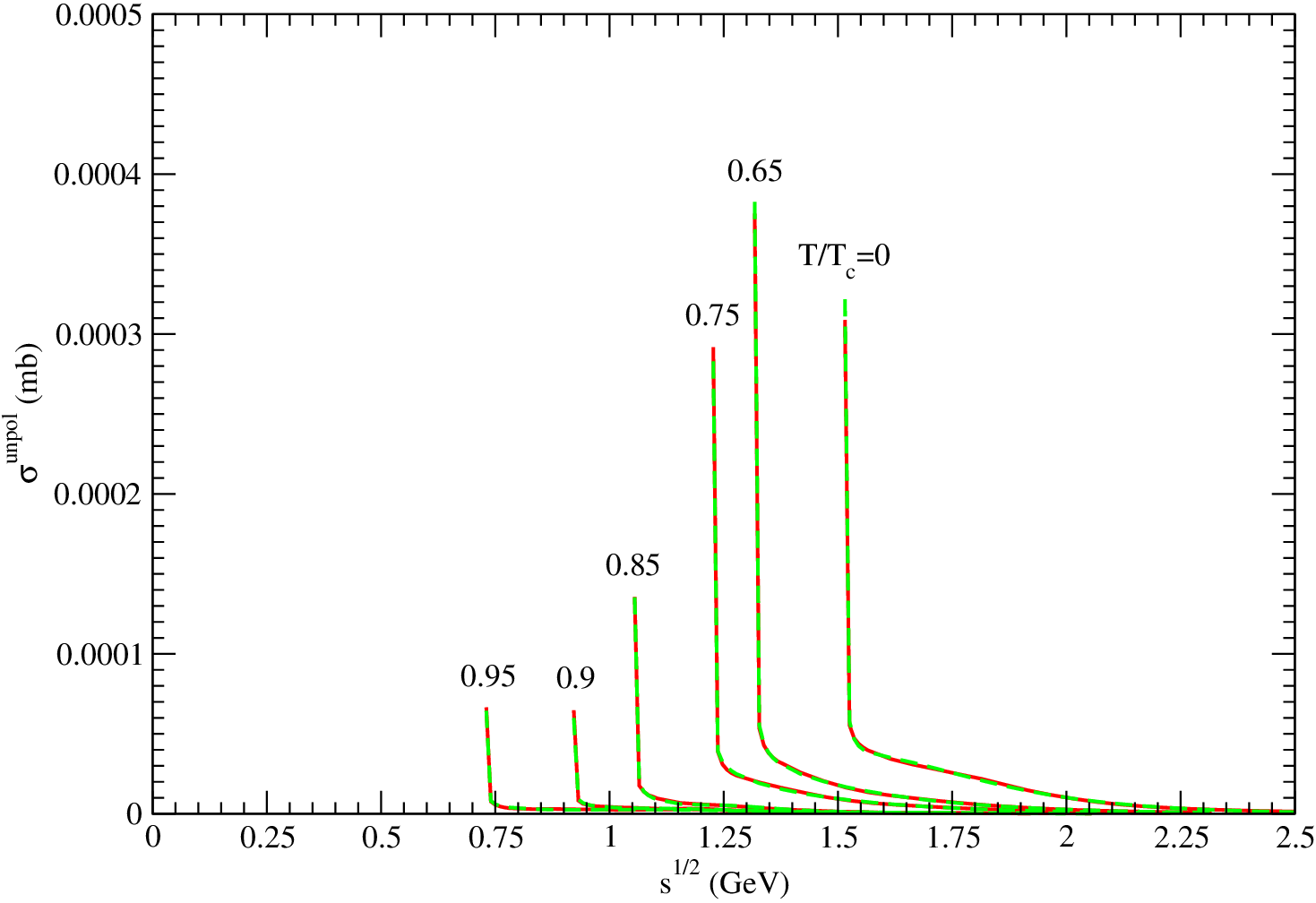}
\caption{Cross sections for $K \phi \to \pi K$ at various temperatures. Red 
solid curves and green dashed curves are obtained from Eq. (35) and Eq. (37),
respectively.}
\label{fig1}
\end{figure}

\begin{figure}[htbp]
\centering
\includegraphics[scale=0.65]{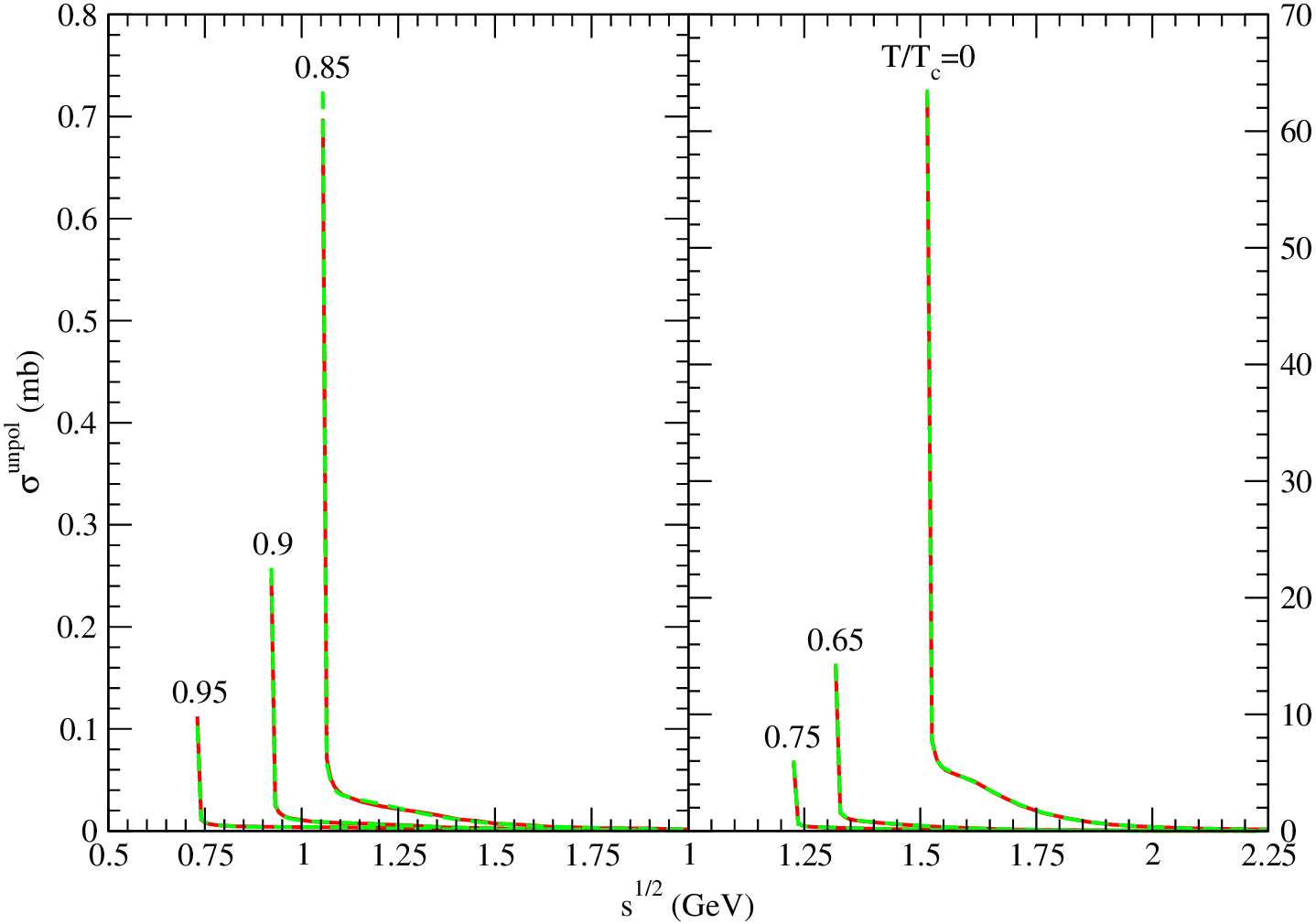}
\caption{The same as Fig. 1 except for $K \phi \to \rho K$.}
\label{fig2}
\end{figure}

\begin{figure}[htbp]
\centering
\includegraphics[scale=0.65]{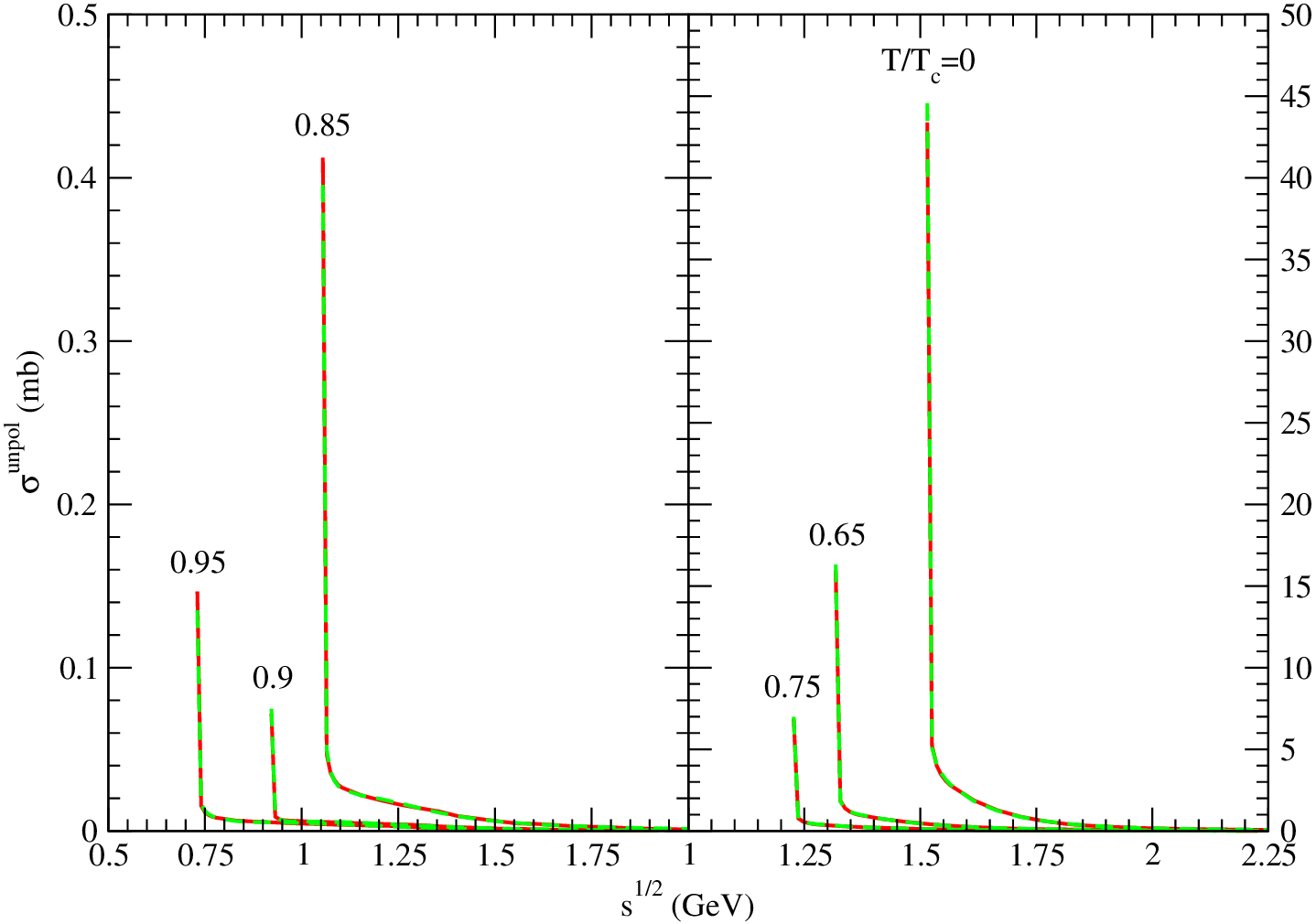}
\caption{The same as Fig. 1 except for $K \phi \to \pi K^*$.}
\label{fig3}
\end{figure}

\begin{figure}[htbp]
\centering
\includegraphics[scale=0.65]{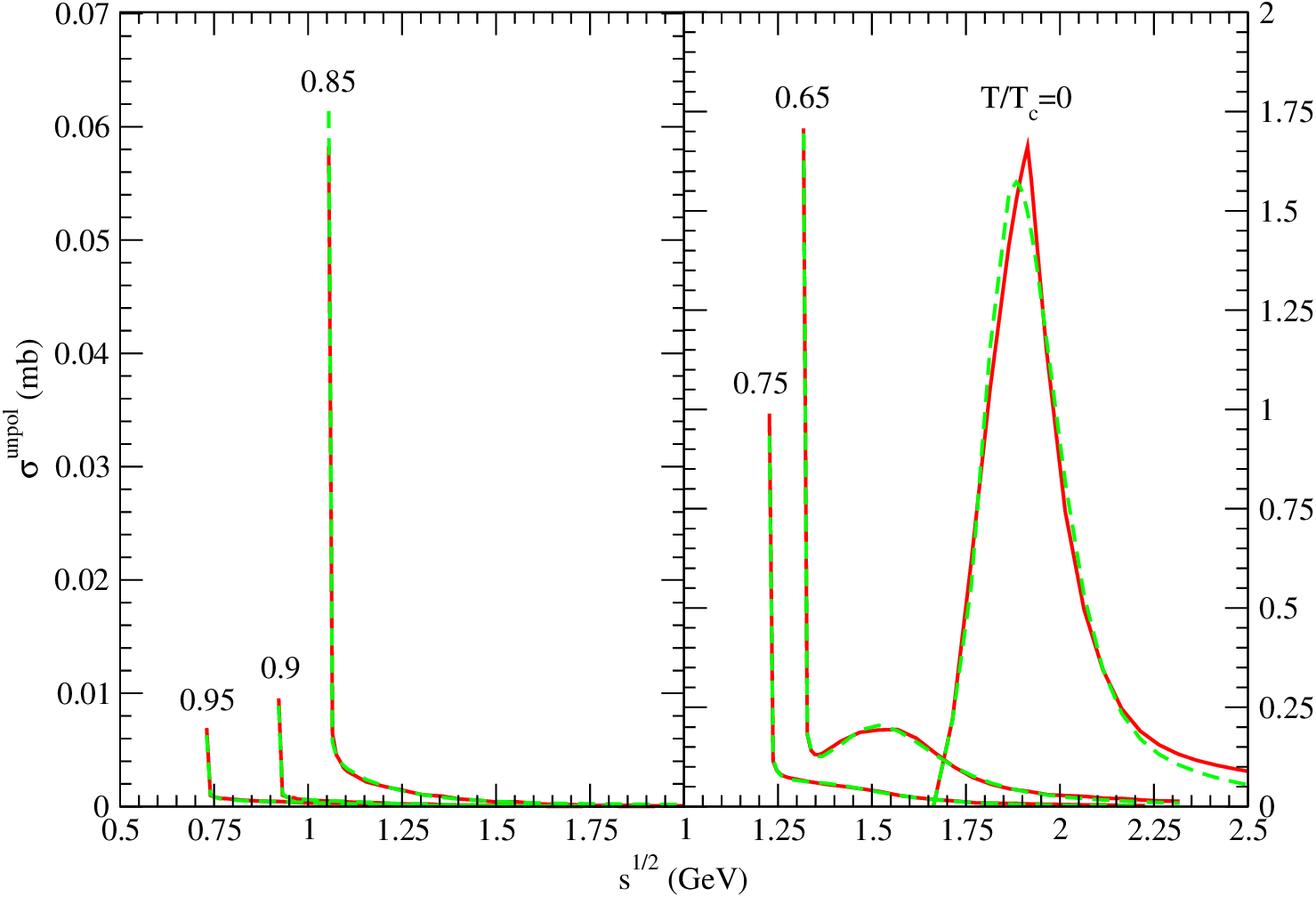}
\caption{Cross sections for $K \phi \to \rho K^*$ at various temperatures. Red 
solid curves and green dashed curves are obtained from Eq. (35) and Eqs. 
(36)-(37), respectively.}
\label{fig4}
\end{figure}

\begin{table*}[htbp] 
\centering\caption{Values of the parameters. $a_1$ and $a_2$ are in units of 
millibarns; $b_1$, $b_2$, $d_0$, and $\sqrt{s_{z}}$ are in units of GeV; $c_1$ 
and $c_2$ are dimensionless.}
\tabcolsep=5pt
\begin{tabular}{*{12}{c}}
\hline
reaction & $T/T_{c}$ & $a_{1}$ & $b_{1}$ & $c_{1}$ & $a_{2}$ & $b_{2}$ & 
$c_{2}$ & $d_{0}$ & $\sqrt{s_{z}}$ \\
\hline
$K\phi\to\pi K$
& 0 & 0.0000026 & 0.17 & 0.62 & 0.0000046 & 0.27 & 1.95 & 0.3 & 2.83 \\ 
& 0.65 & 0.000002 & 0.13 & 0.54 & 0.000003 & 0.24 & 1.2 & 0.15 & 2.57 \\
& 0.75 & 0.0000002 & 0.1 & 0.25 & 0.0000037 & 0.19 & 0.86 & 0.2 & 2.46 \\
& 0.85 & 0.00000076 & 0.239 &  8.37 & 0.00000124 & 0.136 & 0.58 & 0.2 & 2.15\\
& 0.9 & 0.00000042 &  0.271 & 4.7 & 0.00000094 & 0.237 & 0.59 & 0.3 & 1.97 \\
& 0.95 & 0.0000008 & 0.43 & 5.9 & 0.0000009 & 0.23 & 0.55 & 0.45 & 2.17 \\
$K\phi\to\rho K$
& 0 & 0.64 & 0.125 & 2.21 & 0.6 & 0.107 & 0.53 & 0.1 &  2.87 \\
& 0.65 & 0.11 & 0.12 & 0.52 & 0.07 & 0.14 & 2.15 & 0.15 &  2.65 \\
& 0.75 & 0.04 & 0.137 & 0.49 & 0.03 & 0.148 & 1.93 & 0.15 & 1.4 \\
& 0.85 & 0.004 & 0.183 & 2.54 & 0.005 & 0.167 & 0.48 & 0.25 & 2.81 \\
& 0.9 & 0.00233 & 0.202 & 0.5 & 0.00115 & 0.2582 & 4.2 & 0.25 & 2.2 \\
& 0.95 & 0.0009 & 0.371 & 6.63 & 0.0014 & 0.258 & 0.53 & 0.4 & 3.52 \\
$K\phi\to\pi K^*$ 
& 0 & 0.1 & 0.153 & 0.34 & 0.33 & 0.096 & 0.95 & 0.1 & 2.94 \\ 
& 0.65 & 0.121 & 0.132 & 0.51 & 0.068 & 0.121 & 1.84 & 0.15 & 2.69 \\
& 0.75 & 0.05 & 0.147 & 0.5 & 0.027 & 0.127 & 1.65 & 0.15 & 2.52 \\
& 0.85 & 0.0025 & 0.2 & 3.4 & 0.0044 & 0.18 & 0.55 & 0.2 &  2.14 \\
& 0.9 & 0.00089 & 0.211 & 0.54 & 0.00145 & 0.256 & 2.24 & 0.35 & 2.27 \\
& 0.95 & 0.0014 & 0.358 & 3.9 & 0.0016 & 0.153 & 0.55 & 0.3 & 1.86 \\
$K\phi\to\rho K^*$
& 0 & 0.29 & 0.17 & 0.72 & 1.29 & 0.22 & 6.18 & 0.25 & 3.04 \\
& 0.65 & 0.062 & 0.14 & 0.49 & 0.096 & 0.226 & 3.9 & 0.25 & 4.19 \\
& 0.75 & 0.013 & 0.202 & 3.04 & 0.019 & 0.123 & 0.55 & 0.2 & 2.48 \\
& 0.85 & 0.00018 & 0.34 & 0.28 & 0.00059 & 0.11 & 0.91 & 0.15 & 2.16 \\
& 0.9 & 0.000087 & 0.199 & 1.86 & 0.000105 & 0.153 & 0.54 & 0.25 & 1.48 \\
& 0.95 & 0.00007 & 0.26 & 2.7 & 0.00013 & 0.17 & 0.62 & 0.2 & 18.02 \\
\hline
\end{tabular}
\end{table*}

In hadronic matter with high temperatures the four reactions considered in the
present work are exothermic, and always take place. $K$ and $\phi$ mesons
satisfy the Bose-Einstein distribution functions. The reactions and the 
distribution reduce the $\phi$ number and affect the $\phi$ momentum spectra
and the $\phi$ nuclear modification factor, which are
observed in ultrarelativistic heavy-ion collisions 
\cite{STAR2007,PHENIX2011,ALICE2017}.

The total spin of the two initial mesons in the reaction $K\phi\to\pi K$ is 1,
and the total spin of the two final mesons is 0. Because the two total spins 
are unequal, the cross section for $K\phi\to\pi K$ is very small. About
$K\phi\to\rho K$, $K\phi\to\pi K^*$, and $K\phi\to\rho K^*$ the total spin of
the two initial mesons may equal the one of the two final mesons, and the cross
sections may be a few millibarns when $\sqrt s$ is not at the threshold energy.

\vspace{0.5cm}
\leftline{\bf IV. SUMMARY }
\vspace{0.5cm}

With the development in spherical harmonics of the relative-motion wave 
functions of the two initial mesons and of the two final mesons, new 
expressions
of the transition amplitudes have been obtained. With the transition amplitudes
we have
calculated unpolarized cross sections for $K\phi\to\pi K$, 
$K\phi\to\rho K$, $K\phi\to\pi K^*$, and $K\phi\to\rho K^*$, which are governed
by quark-antiquark annihilation and creation. Both parity conservation and
total-angular-momentum conservation are maintained. To use the numerical 
cross sections conveniently, we have parametrized the cross 
sections. Each of the exothermic reactions $K\phi\to\pi K$, $K\phi\to\rho K$, 
and $K\phi\to\pi K^*$ exhibits a rapid decrease first and then a slow decrease
in cross section with 
increasing $\sqrt{s}$ from the threshold energy. That
the reaction $K\phi\to\rho K^*$ is endothermic or exothermic depends on
temperature. The temperature-dependent cross sections are related to 
temperature dependence of the quark potential, the quark-antiquark 
relative-motion wave functions, and the meson masses.

\vspace{0.5cm}
\leftline{\bf ACKNOWLEDGEMENTS}
\vspace{0.5cm}

This work was supported by the National Natural Science Foundation of China
under Grant No. 11175111.

\end{document}